\documentclass[preprint,pra,showpacs]{revtex4}   

\usepackage{graphicx} 
\usepackage{dcolumn}  
\usepackage{bm}       

\begin{document}

\preprint{.}

\title{Two-electron quantum dots as scalable qubits}

\author{J.H. Jefferson}
 \email{jhjefferson@qinetiq.com}
\author{M. Fearn}
\author{D.L.J. Tipton}
\affiliation{QinetiQ, Sensors and Electronics Division, St Andrews Road, Malvern, WR14 3PS, UK}

\author{T.P. Spiller}
\affiliation{Hewlett-Packard Laboratories, Filton Road, Stoke Gifford, Bristol, BS34 8QZ, UK}

\date{\today}

\begin{abstract}
We show that two electrons confined in a square semiconductor quantum dot
have two isolated low-lying energy eigenstates, which have the potential to
form the basis of scalable computing elements (qubits). Initialisation, one-qubit
and two-qubit universal gates, and readout are performed using electrostatic gates
and magnetic fields. Two-qubit transformations are performed via the Coulomb 
interaction between electrons on adjacent dots. Choice of initial states and 
subsequent asymmetric tuning of the tunnelling energy parameters on adjacent
dots control the effect of this interaction.
\end{abstract}

\pacs{03.67.Lx, 68.65.+g, 73.23.Ad, 73.20.Dx, 71.45.Gm, 73.61.-r}

\maketitle

\section{Introduction}

   Interest in quantum computing began to grow rapidly after the discovery
of the factoring \cite{Shor94,Shor97} and searching \cite{Grover96} algorithms,
which demonstrate that if a quantum computer could be built it would be capable
of tasks impossible with conventional classical machines. This interest was 
further stimulated by the development of quantum error correction \cite{Shor95,Steane96} and 
fault tolerant \cite{Steane97,Shor96} techniques. These show that, rather than
requiring the decoherence of the qubits comprising a quantum computer to be 
completely negligible during the evolution corresponding to a computation,
it simply needs to be small. Then the redundant encoding of quantum information 
using additional quantum bits (and the more sophisticated fault-tolerant 
techniques) can be employed to maintain the coherence of the qubits effectively 
responsible for running the quantum algorithm. All this was a real stimulation 
for quantum computing hardware. Real quantum systems always have some level of 
decoherence and whilst it is hard to envisage removing this completely, making 
it small is a reasonable practical goal.

    Consequently, over the past few years numerous systems have been examined
(theoretically and/or experimentally) to assess their potential for qubits and
quantum computing hardware \cite{Forschritte00}. These include photons \cite{Knill01}, ions in an
electromagnetic trap \cite{Cirac95,Cirac00,Monroe95,Sackett00}, atoms in beams 
interacting with cavities at optical \cite{Kimble77} or microwave \cite{Haroche97}
frequencies, electronic \cite{Ekert96} and spin \cite{Loss98} states in quantum dots,
nuclear spins in a molecule in solution \cite{Gershenfeld97,Cory97} or in solid 
state \cite{Kane98}, charge (single Cooper pair) states of nanometre-scale 
superconductors \cite{Shnirman97,Nakamura99}, flux states of
superconducting circuits \cite{Bocko97,Mooij99,Friedman00,VanDerWal00}, quantum
Hall systems \cite{Privman98} and states of electrons on superfluid helium \cite{Platzman99}.

    Five main criteria have been identified for quantum computing hardware \cite{Forschritte00,DiVincenzo97}:
(i) Clearly identifiable qubits and the ability to \textit{scale up} in qubit number;
(ii) The ability to prepare ``cold'' initial states (such as the thermal ground state
of the whole system); (iii) Small decoherence (so correction \cite{Shor95,Steane96,Steane97} can be utilized) 
- a rough benchmark is a fidelity loss of $10^{-4}$ in an elementary quantum 
gate operation; (iv) The ability to realize a universal set of quantum gates -- arbitrary
single qubit rotations and an entangling two-qubit gate are enough to build up any 
general evolution \cite{DiVincenzo95,Lloyd95,Barenco95}; (v) The ability to measure 
(in the usual sense of projective quantum measurements) qubits to determine the
outcome of a computation. 

     Any qubit candidates have to be assessed against this ``DiVincenzo checklist'', 
to see how they fare. Although all the above-mentioned candidates do reasonably well, 
there is to date no clear favourite route for quantum computing hardware. It is certainly
true that so far the few-qubit experiments which have been performed have used fundamental
entities for qubits, such as photons, atoms, ions or nuclear spins (in liquid state NMR).
Such systems exist or can be placed in low-decoherence environments and have thus 
provided the first few-qubit demonstrations. However, it seems likely that \textit{scalabilty}
in qubit number, up to the size of useful many-qubit devices, may well be more easily 
achieved with condensed matter systems, making use of fabrication or self-assembly techniques
to produce many-qubit arrays. The trade-off is that decoherence in such systems is generally
higher than for fundamental qubits. 

     At present, there is thus considerable interest in solid state approaches to qubits.
These have to match up well against (i), (ii), (iv) and (v) of DiVincenzo's list, and in 
the first instance for (iii) must at least offer small enough decoherence for the demonstration 
of quantum gates. With flexibility in fabrication, choice of materials, etc., it is then 
possible that decoherence may be reduced down to the required level for actual quantum 
computation. In this paper we discuss a solid-state approach to qubits based on a particular 
form of quantum dot.

\begin{figure}
\includegraphics[width=4in]{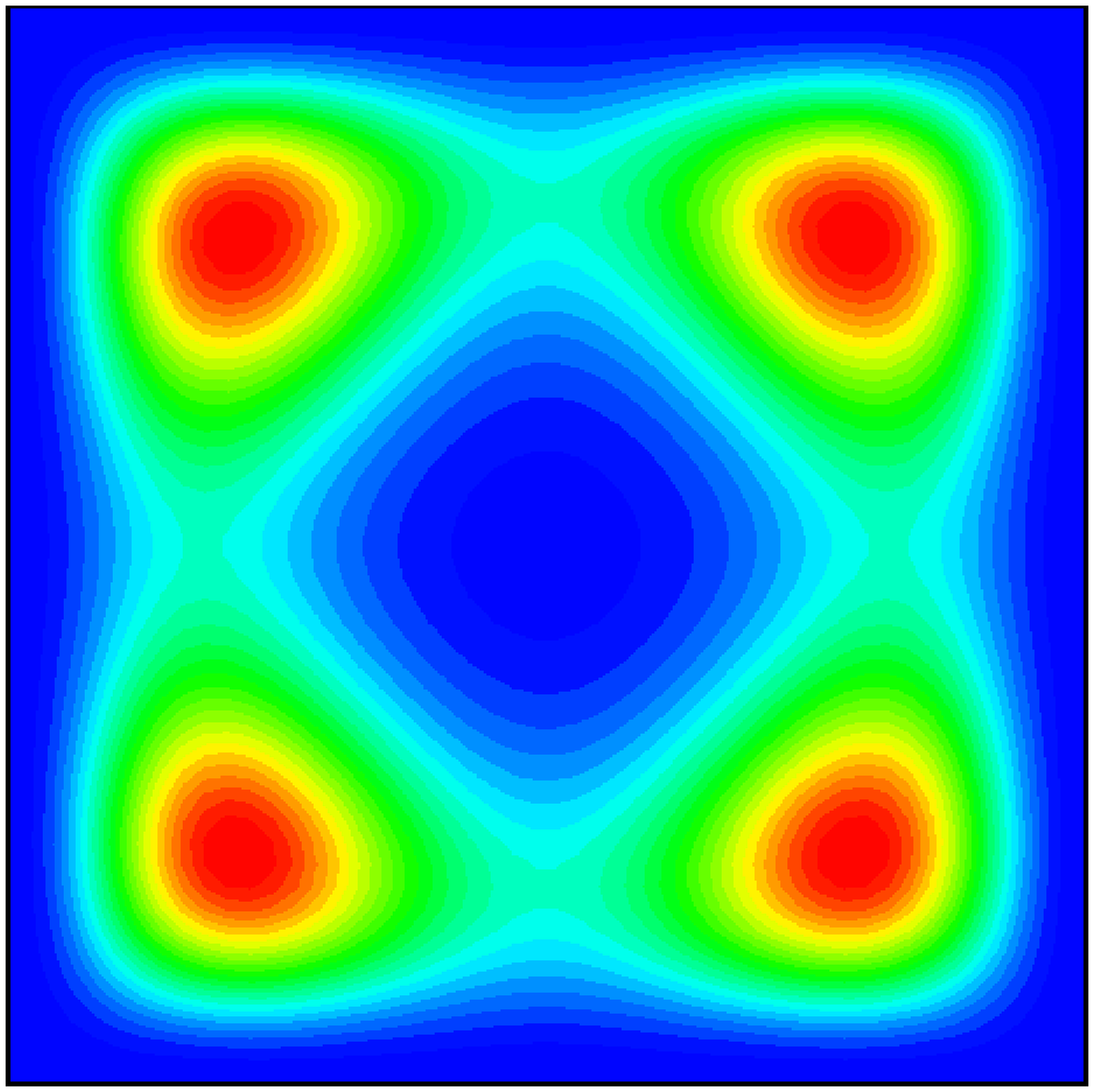}
\caption{\label{fig1}Ground-state charge density, peaked in diagonally-opposite corners, for a two-electron GaAs quantum dot with side length 800nm ($\approx$100 effective Bohr radii).}
\end{figure}

     In some recent work \cite{Jefferson97,Creffield99} we have shown that a few 
electrons in polygonal quantum dots, which are sufficiently large that the mean 
electron separation is greater than their  effective Bohr radius, always have a 
charge-spin multiplet at low-energy which is well separated 
from higher-lying states. This isolation is a consequence of electron correlations, which 
tend to force the electrons into configurations close to their classical electrostatic 
ground state.  Such states always have a quasi-degeneracy of  at least $2^N$
where $N$ is the number of electrons, and further degeneracies  may arise due to 
equivalent electrostatic minima \cite{Jefferson97}.  An isolated ground multiplet 
of degeneracy $m$ suggests that such systems could form the basis for a realisation 
of quantum computing  elements, where each element is an $m$ state quantum system 
or `qumit'. In this paper we focus on one such possibility, a square 2D semiconductor 
quantum dot containing two electrons. When the dot is large compared with the electron 
effective Bohr radius, the Coulomb repulsion dominates and tends to repel the electrons 
to diagonally opposite corners of the dot and the charge density is peaked in these 
regions, as shown in figure \ref{fig1}, in which the exact ground-state charge density is plotted. 
There are in fact 8 states in the ground manifold (2 singlets and 2 triplets) which 
arise from the two base states shown in figure \ref{fig2}, each of which may be a singlet or 
a triplet depending on spin. These states could in principle form a base-8 quantum 
computing element.  However, it is simpler to stay within one spin manifold, which 
is valid provided the spin-flip scattering time is sufficiently long. 
The initial state may be prepared in a given spin state in various ways. For example, in 
zero magnetic field at low temperatures the ground state is a singlet, whereas application 
of a magnetic field will give a triplet lowest when the flux through the dot is of order a 
half-integral number of flux quanta (see next section for further details). The Zeeman splitting 
will then give a spin-polarised ground state.

     In the next section we analyse this system in detail and show how it forms the basis 
for a scalable qubit element. This element is very similar to the so-called quantum cellula 
automata, which has been proposed \cite{Lent93,Bernstein99} as a basic building block for 
a new kind of computing system in which the elements are either in one state (0) or the other (1). 
However, we emphasise that the system proposed here is fundamentally different in that 
we consider cellula  automata as a fully coherent system in which the qubit (or array of 
qubits in general) are in superposition states in the true sense of quantum computation. 
Such quantum coherent cellula automata have been considered recently by T{\'o}th and Lent 
\cite{Toth01} who also suggested square qubit cells consisting of four quantum dots in 
the corners. In this scheme there is vertical tunnelling between dots. Our work 
complements and extends theirs in several ways. Our qubit cells consist of a single 
quantum dot of square geometry in which the electrons are confined by their mutual 
Coulomb repulsion (strong correlation regime). The two-electron states are allowed to 
take on the full square symmetry with electrons tunnelling around the perimeter of the square. 
This lifts the spin degeneracy and produces independent two-state systems for singlets and 
triplets. These states are further controlled by corner gates and external magnetic fields. 
We also propose a new method of producing two-qubit entanglement via the Coulomb interaction . 

\section{Basic Model}

\begin{figure}
\includegraphics{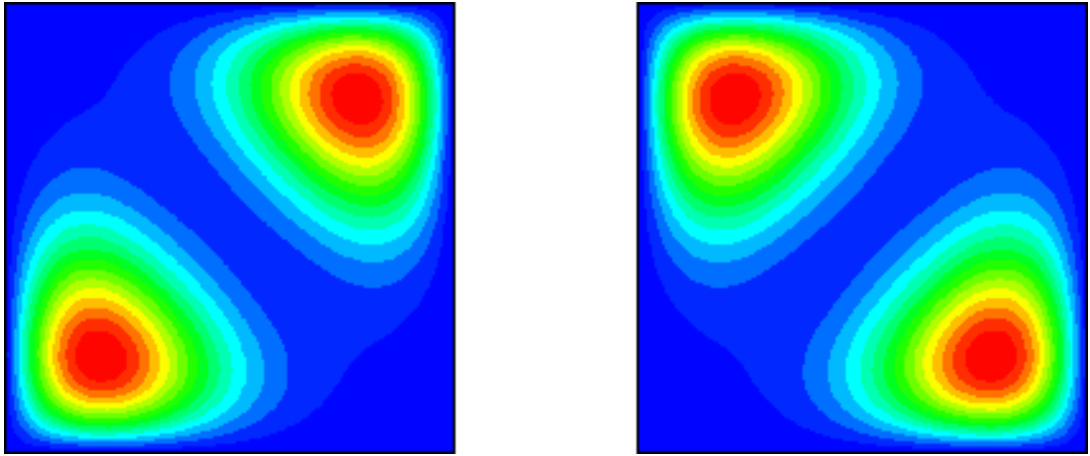}
\caption{\label{fig2}Contour plots of two-electron basis functions which yield the ground-state
charge density of figure \ref{fig1}.}
\end{figure}

\begin{figure}
\includegraphics{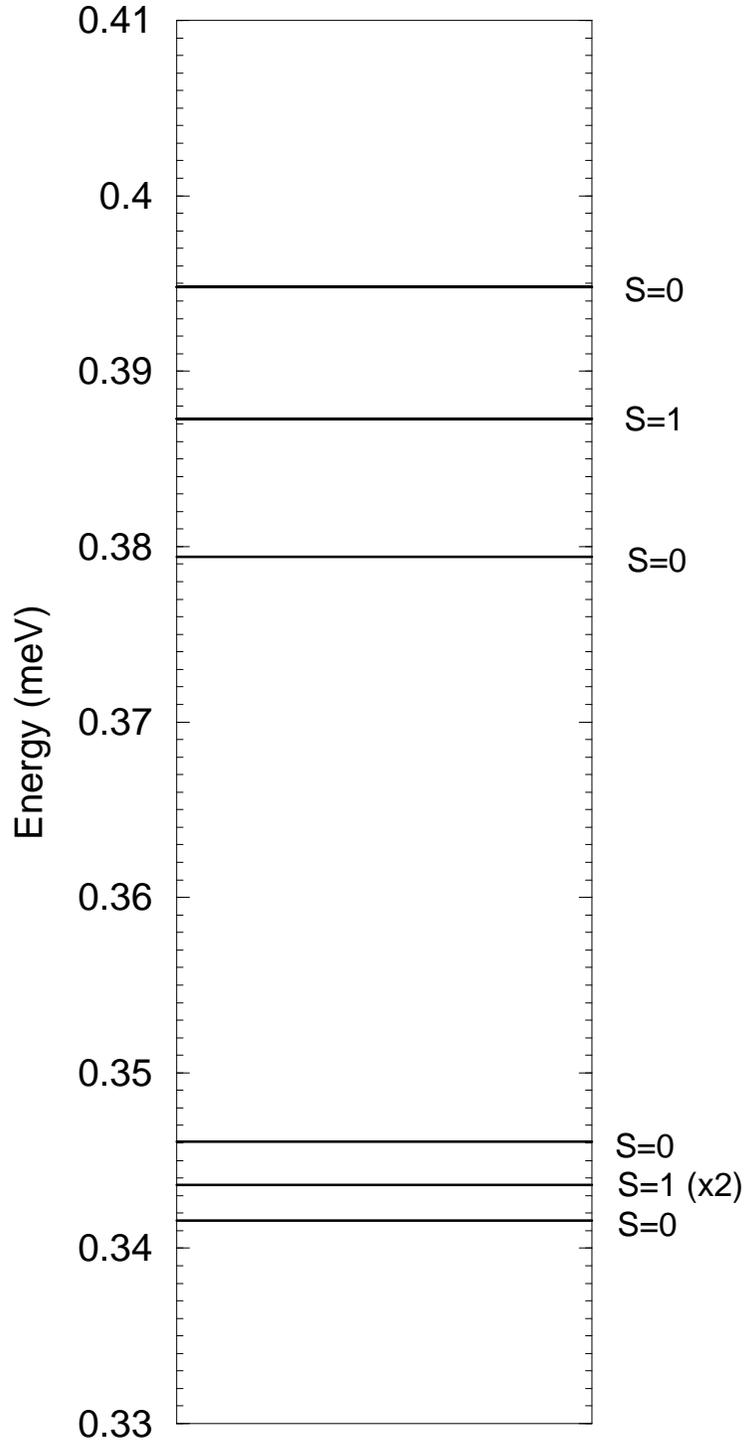}
\caption{\label{fig3}Energy spectrum of a two-electron quantum dot showing an isolated low-lying
manifold of two singlets and two triplets.}
\end{figure}

     We consider two electrons in a semiconductor quantum dot with a square confining 
potential in the $x-y$ plane and sufficiently high confinement in the $z$ direction 
that the electrons always occupy the lowest bound state in this direction. Such structures 
may be fabricated, for example, from a gated two-dimensional electron gas at a heterojunction 
interface. In reality this would not produce a perfectly square confining potential though 
such deviations would not change qualitatively the results which follow and may be accounted 
for quantitatively by a straightforward extension of the theory. Further control of the 
two-electron states may be achieved by placing gate electrodes in the four corners of the dot, 
since applying a positive voltage to a corner electrode would provide a potential well for an 
electron in a region at the heterojunction interface below the gate. We consider low-lying 
two-electron states of the quantum dot in a magnetic field, which may be modelled using the 
2D Schr{\"o}dinger equation (after integrating over $z$) 
\footnote{This is in fact an approximation since the Schr{\"o}dinger equation is not strictly 
separable and integration over the lowest bound-state in the z-direction modifies the 
Coulomb interaction. However, the error due to this approximation is small and unimportant 
for what follows.}

\begin{equation}
i\hbar \frac{{\partial \Psi (x,y,t)}}{{\partial t}} = H\Psi (x,y,t)   \label{eq1}
\end{equation}			
where
\begin{equation}
H = \frac{1}{{2m}}\left[ {\sum\limits_{i = 1}^2 {\left( { - i\hbar \nabla _i  + e{\bf{A}}_i^{} } \right)} ^2  + V({\bf{r}}_i ,t) + E_B s_z^i } \right] + \frac{{e^2 }}{{4\pi \varepsilon |{\bf{r}}_1  - {\bf{r}}_2 |}}   \label{eq2}
\end{equation}
and where $V(\bf{r},t)$ is the total one-electron confining potential, 
$m$ is the effective mass of the electrons and ${\bf A}$ is the vector potential 
of a magnetic field ${\bf B}$ oriented perpendicular to the plane of the dot and 
$E_B=g\mu_B B$ is the Zeeman energy.  When the length-scale of the confining potential
is small compared with the  effective Bohr radius of the electrons, the eigenstates of
$H$ (with time-independent confining potential) may be computed to good accuracy
using the Hartree-Fock approximation. This is rather like what is done in atomic physics
and the term `artificial atoms' has been used to describe such structures \cite{Kastner92}.
However, it should be realised that the energy scale is much smaller than in real atoms
since the effective Bohr radius is typically two orders of magnitude larger. In this paper
we shall be concerned with the opposite limit in which the size of the dot is large
compared with the effective Bohr radius. In this regime, electron correlations become 
dominant since the Coulomb repulsion energy between the electrons scales like $a/L$ whereas
the electron kinetic energy scales like $(a/L)^2$, where $a$ is the effective Bohr Radius 
and $L$ is the confinement length. For large $L/a$ the ground state of the 
electrons tend towards the classical minimum energy given by electrostatics. For the present 
case of two electrons in a square confinement potential, the electrons are pushed to diagonally 
opposite corners, as shown in figure \ref{fig2}. We can see immediately that such a ground state is 
degenerate since there are two equivalent positions which the pair of electrons may occupy 
and furthermore their spins may each take two values, giving a total degeneracy of 8 in the 
limit of infinitely large dots. For finite sized dots with $L/a>>1$, kinetic energy 
partly lifts this degeneracy resulting in two singlets, with two degenerate triplets lying 
approximately midway between the singlets in zero magnetic field. This spectrum is shown 
in figure \ref{fig3} for the case of a square quantum dot of side length $800 nm$ and material parameters 
corresponding to GaAs, giving a ratio $L/a \approx 100$. 
Also shown in the spectrum are the 
next few excited states, which are somewhat higher in energy. In fact this isolation of a 
low-lying multiplet is a common feature of few-electron states in the strongly correlated 
limit to which we refer to reference \onlinecite{Jefferson97} for further details. This feature 
was also one of  our main motivations for considering their use as quantum computing elements.
However, one  price to pay for this is that the absolute energy scale is small, though this is material 
dependent. For the given example, the singlet-separation in the ground manifold is of order $5\mu eV$
whereas the next singlet is some $33\mu eV$ higher in energy, as shown in figure \ref{fig3}. This spectrum 
was obtained numerically by expanding the eigenstates in basis functions of the noninteracting 
electron system and diagonalising the resulting Hamiltonian matrix. We emphasise that this 
seemingly simple problem requires considerable computing resources to obtain convergence 
for the lowlying eigenstates when the dot is large. For example, in the case of the $800 nm$ 
dot, the order of the matrices to be diagonalised was around $10000$ and this grows rapidly 
with dot size. However, the simple electrostatic picture suggests that for the lowlying 
multiplet a simpler approximate method should be possible in which the electrons are in 
localised states near the corners of the dot. We have shown that this is the case in which 
one-electron Hartree states may be constructed, in which the electron is in one of the 
four corners. These are then used to construct suitable anti-symmetrised two-electron states, 
rather like Heitler-London molecular states. Restricting ourselves to the lowest Hartree 
states in the `ground manifold' and eliminating higher energy states by quasi-degenerate 
perturbation theory gives excellent agreement with the numerically exact results and is 
computationally very efficient \cite{Creffield00}. More importantly, it gives us a simple
description  of the lowest multiplet shown in figure \ref{fig3}, which may be described by the
following effective Hamiltonian :
\begin{equation}
H_{{\rm{eff}}}  = \sum\nolimits_{i = 1}^4 {\varepsilon _i (B)} n_i  + \left( {\Delta (B)e^{i2\phi } R_{\pi /2}  + {\rm{H.c.}}} \right)  \label{eq3}
\end{equation}
where $\varepsilon_i(B) = \varepsilon_i(0) + E_B s_z^i$, 
$n_i = c^\dagger_{i \uparrow} c_{i \uparrow} + c^\dagger_{i \downarrow} c_{i \downarrow}$,
$R_{\pi/2}$ is a rotation operator which rotates both electrons simultaneously through 
an angle $\pi/2$ about a point at the centre of the dot, $\Delta(B)$
is an energy parameter which has only weak dependence on $B$ (which we neglect)
and $\phi$ is a Pierls phase factor which an electron picks up when hopping from a localised 
(Hartree) state on one corner to an adjacent corner. This phase factor is related to 
the total magnetic flux through the dot by $4\phi = 2 \pi \Phi/\Phi_0$, where $\Phi$ is 
the total flux through the dot and $\Phi_0 = h/e$ is the flux quantum
\footnote{In fact $\Phi=B A$, where $A$ is an effective area of the dot which is 
slightly smaller than its actual area, reflecting the fact that the charge density 
peaks for the electrons are not precisely in the corners of the dot.}. The zero field 
one-electron energies, $\varepsilon_i(0)$, depend on the corner gate voltages and 
may be used to force the ground-state of the system to be either one or the other of 
the states shown in figure \ref{fig2}. Conversely, when all 4 corner gate voltages are the 
same, the ground state is then an equal superposition of these base states. It is 
important to realise that $H_{\rm eff}$ only operates in the subspace 
defined by these base states, all other states consisting of electrons occupying 
adjacent corners of the dot must be omitted from the basis set. All such `excited' 
base states (and many others corresponding to higher energy Hartree states) are 
accounted for in the energy parameters of equation (\ref{eq3}). We stress that these parameters 
are calculated from the underlying Hartree states, not fitted to the numerically 
exact solutions. The method thus gives an efficient scheme for analysing the low-lying 
multiplet and calculating the associated energy parameters and their variation with 
magnetic field, dot geometry and gate potentials. 

    It is now straightforward to diagonalise $H_{\rm eff}$ within this restricted 
subspace for fixed magnetic field and gate potentials. If we neglect the Zeeman term, 
then the eigenstates must consist of two singlets and two triplets. These correspond 
to symmetric and anti-symmetric orbital states respectively. It is straightforward to 
show that the resulting $2X2$ Hamiltonian matrices are:
\begin{equation}
H_{{\rm{singlet}}}  = \left[ {\begin{array}{*{20}c}
   {E_0 (B)} & {2\Delta \cos (2\phi )}  \\
   {2\Delta \cos (2\phi )} & {E_1 (B)}  \\
\end{array}} \right]
\label{Hsing}
\end{equation}
and
\begin{equation}
H_{{\rm{triplet}}}  = \left[ {\begin{array}{*{20}c}
   {E_0 (B)} & {2\Delta \sin (2\phi )}  \\
   {2\Delta \sin (2\phi )} & {E_1 (B)}  \\
\end{array}} \right]
\label{Htrip}
\end{equation}
where $E_0(B)=\varepsilon_1(B)+\varepsilon_3(B)$ and $E_1(B)=\varepsilon_2(B)+\varepsilon_4(B)$,
numbering clockwise from the top left-hand corner. The eigenenergies are thus,
\begin{equation}
E_{{\rm{singlets}}}  = \frac{{E_0  + E_1  \pm \sqrt {(E_0  - E_1 )^2  + 16\Delta ^2 \cos ^2 2\phi } }}{2}
\end{equation}
and
\begin{equation}
E_{{\rm{triplets}}}  = \frac{{E_0  + E_1  \pm \sqrt {(E_0  - E_1 )^2  + 16\Delta ^2 \sin ^2 2\phi } }}{2}
\end{equation}

\begin{figure}
\includegraphics[height=4in]{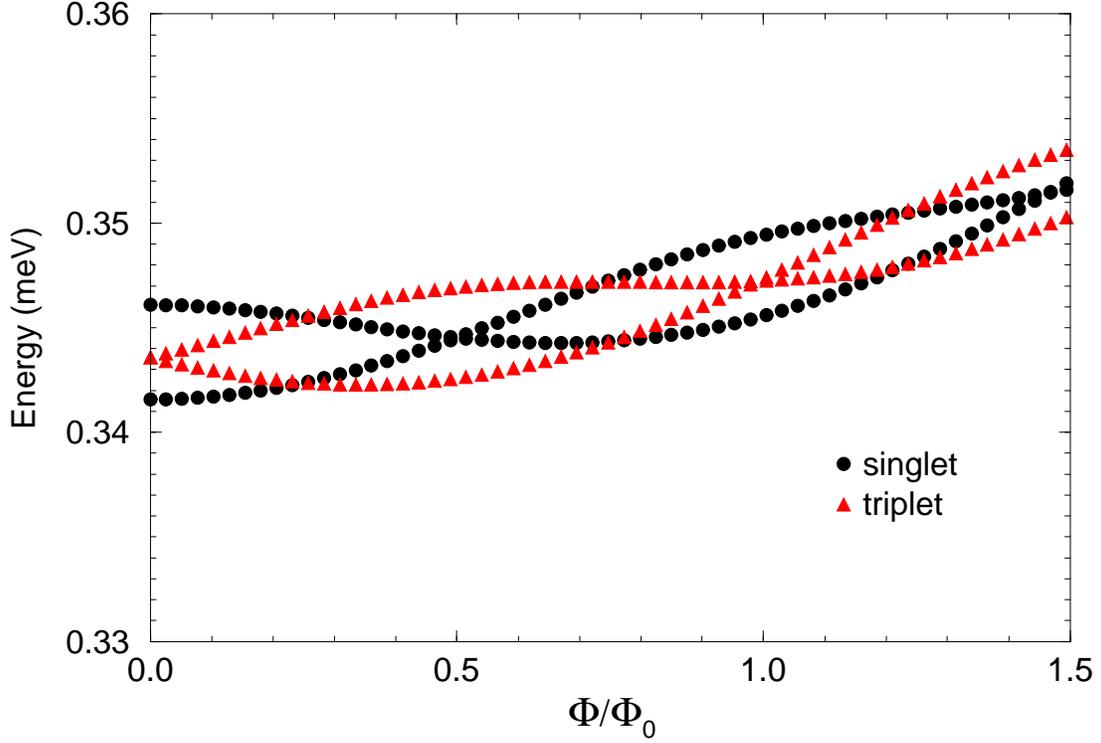}
\caption{\label{fig4}Variation of low-lying eigenenergies with magnetic flux of a two-electron
quantum dot. $\Phi_0 = h/e$ is the flux quantum.}
\end{figure}

These energies are in agreement with the exact low-lying multiplet shown in 
figure \ref{fig3} for the special case when all corner gate voltages are the same 
($E_0 = E_1$) and zero magnetic field ($\phi=0$).  At finite magnetic field, the energy 
of the singlets and triplets oscillate sinusoidally and also increase overall 
in energy with increasing magnetic flux through the dot, as shown in figure \ref{fig4}. 
The important points to note are that (i) the ground multiplet becomes a triplet 
with maximum splitting to the other triplet when half a flux quantum passes 
through the dot (and also for odd multiples of half a flux quantum) (ii) the 
coupling between states of the same symmetry may be `tuned' by the magnetic 
field, e.g. when half a flux quantum passes through there is no coupling between 
singlets (degenerate) and maximum coupling between triplets \footnote{In fact the
role of singlets and triplets reverse for half flux quantum through the dot. It is
straightforward to show that the electrons have the same behaviour as fictitious
hard core spin-bosons. The `particles' may be regarded as `composite fermions' for
this two-electron system.}. These properties 
may be exploited for quantum computing in that the interstate coupling (and 
hence time scale) may be tuned. Furthermore, when the Zeeman term is switched 
on, the triplets split with the $S_z=-1$ states lowest. 
Thus we may initialise the system to be in the lowest spin-polarised 
$S_z=-1$ by choosing the magnetic field so that the magnetic 
flux through the dot is a half-integral, allowing the system to settle into its ground 
state at low temperature ($T<<E_B$). Alternatively, we may 
make a spin-singlet lowest in energy by choosing a magnetic field for which an integral 
number of flux quantum pass through the square, or indeed zero magnetic field. The 
choice of singlets or spin-polarised triplets for the qubit states is to some extent 
arbitrary but the choice should take into account the energy scales of the Zeeman 
spiltting and the singlet-triplet splitting. For large dots, $E_B$
can easily exceed $\Delta$ and the spin-polarised state is preferred since all 
other states are pushed to higher energies. This has the disadvantage that the 
absolute energy/temperature scale is low, e.g. for an $800nm$ square dot
$\Delta\approx1\mu eV=12 mK$ in GaAs for which the Zeeman energy 
for a magnetic field of 1 tesla is about $25 \mu eV=0.3 K$. 
Although such large dots have obvious advantages for fabrication and lithography, 
they would require very clean samples. Conversely, decreasing the dot size to, say, 
100nm raises $\Delta$ to around $0.5meV=6K$ which is now somewhat greater than the 
1 tesla Zeeman splitting. With corner gates this would be close to the present limits of what can 
be achieved by electron-beam lithography.  

As with any 2-state system, we can map the Hamiltonian onto a pseudo-spin model, 
which takes the form,

\begin{equation}
H_{ps}  = \bar E + \varepsilon \sigma _z  + \gamma \sigma _x 
\label{spinH}
\end{equation}
where $\sigma_x$ and $\sigma_z$ are the usual Pauli operators,  
$\bar E = (E_0 + E_1)/2, \varepsilon=(E_1-E_0)/2, \gamma=2\Delta \cos(2\phi)$ for singlets 
and $\gamma=2\Delta \sin(2 \phi)$ for triplets. 
Despite the simplicity of this effective Hamiltonian, the main physics of the 
two-electron qubit system is contained within the energy parameters in equation (\ref{spinH}), 
which have, in general, complicated functional dependencies on the confining region 
(defined by the boundary of the dot and gate electrode potentials), the mutual 
coulomb repulsion of electrons within the dot, and the external magnetic field 
through the dot. Nevertheless, these dependencies may be computed for any specific 
geometry and magnitudes of external potentials and fields and hence the parameters 
calibrated.  In what follows we shall continue to refer to the 2-electron 
pseudo-spin base states as $|0\rangle$ and $|1\rangle$ to distinguish them from 
true single-electron spin base states $|\uparrow\rangle$ and $|\downarrow\rangle$.

    The Hamiltonian (\ref{eq2}) and corresponding effective Hamiltonians (\ref{eq3})
and (\ref{spinH}) are readily generalised to two or more identical dots each containing two electrons. 
Such structures could be fabricated with gates and we envisage that the separation 
between the dots would be greater than the size of the dots themselves. This ensures 
that the Coulomb repulsion between electrons on different dots is always smaller 
than the Coulomb repulsion between the two electrons on the same dot and may thus 
be treated as a perturbation. 
Furthermore, the confining barriers and dot-dot separation ensures that tunnelling 
of electrons between dots, even as virtual states, is negligible.  In this situation, 
only diagonal elements of the Coulomb interaction of the density-density type are 
non-negligible.  Hence, the effective Hamiltonian for an array of $N$ 
two-electron quantum dots may be written, in pseudo-spin notation (dropping a 
constant energy term):
\begin{equation}
H_{ps} = \sum\limits_{i = 1} {\left[ {\varepsilon _i \sigma _{iz}  + \gamma _i \sigma _{ix}} \right]} - \frac{1}{2}
\sum\limits_{i ,j=1; i \ne j}{v_{ij} \sigma _{iz} \sigma _{jz} } 
\label{effspinH}
\end{equation}
where the intra-dot energy parameters have the same meaning as in equation (\ref{spinH}) 
and $v_{ij}$ is half the difference in coulomb repulsion energy between dots $i$ and 
$i$ with parallel and anti-parallel pseudo-spins, i.e.
\begin{equation}
v_{ij} = \frac{\langle 0_i, 1_j | V_2 | 0_i, 1_j \rangle -
            \langle 0_i, 0_j | V_2 | 0_i, 0_j \rangle }{2}
\end{equation}
where
\begin{equation}
V_2 = \frac{e^2}{4 \pi \epsilon |{\bf r_1} - {\bf r_2}|}.
\end{equation}
Note that by symmetry, $\langle 00 | V_2 |00 \rangle = \langle 11 | V_2 |11 \rangle$ and
$\langle 00 | V_2 | 01 \rangle =  \langle 00 | V_2 | 10 \rangle =  \langle 11 | V_2 | 01 \rangle = \langle 11 | V_2 | 10 \rangle$.

If we neglect screening between electrons on different dots then since
the electron wavefunctions are localised in the corner of the dots,
\begin{eqnarray}
v_{n} & \equiv  & v_{i,i+n}\simeq \frac{e^{2}}{8\pi \varepsilon }
\Big(\frac{1}{nd-L}+\frac{2}{\sqrt{(nd)^{2}+L^{2}}}+\frac{1}{nd+L}- \\ \nonumber
      &         & \frac{2}{nd}-\frac{1}{\sqrt{(nd-L)^{2}+L^{2}}}-
		   \frac{1}{\sqrt{(nd+L)^{2}+L^{2}}}\Big)
\end{eqnarray}
where $d$ is the spacing between adjacent dots. Expanding (1) in
$\frac{L}{nd}$ to lowest-order gives,
\[
v_{n}\simeq \frac{3e^{2}}{4\pi \varepsilon nd}\Big(\frac{L}{nd}\Big)^{4}
\]
In practise, there will be some screening due to image charges in
the corner gates giving,
\[
v_{n}\simeq \frac{5e^{2}}{2\pi \varepsilon }\Big(\frac{\delta }{nd}\Big)^{2}\Big(\frac{L}{nd}\Big)^{4}
\]
where $\delta$ is the distance between an electron and its image
charge. Hence, $\frac{v_{n}}{v_{1}}=n^{-6}$ and while this decay is
quite rapid, it is still sufficiently slow that the neglect of second
and higher neighbour interactions would give rise to significant errors \cite{Referee01}.
In section IV we show how such systematic errors may be reduced to 
acceptable levels by changing the tunnelling parameter $\gamma $
on some of the dots. This effectively turns off the qubit interaction.
In addition to restricting the effect of the Coulomb interaction to
nearest neighbours, this technique is also used when performing single-qubit
transformations for which even the nearest-neigbour coupling is turned
off, as discussed elsewhere for qubits based on Josephson junctions 
\cite{Schon99,Schon00}.

In the next two sections we will show how 2-electron square quantum dots described by 
this model and the pseudo-spin Hamiltonian (\ref{spinH}), have potential for quantum computing elements.

\section{Initialisation and single-qubit gates}

    As emphasised by DiVincenzo and coworkers, a practical quantum computer 
has to satisfy a certain minimal list of requirements \cite{DiVincenzo97}. In this section 
we show how the 2-electron quantum dot can fulfil two of them, namely initialisation 
of the system and a universal single-qubit gate.  Since the 2-electron quantum dot 
is an 8-state system, it is convenient to reduce this to a 2-state `qubit' system.  
This is not strictly necessary as in principle one can deal directly with octal 
elements though this poses many practical problems of implementation.  The simplest 
way to reduce the quantum dot to a 2-state system is to essentially remove the spin 
degrees of freedom, as described in the previous section. This has the added advantage 
that the resulting two charge degrees of freedom may be manipulated electrostatically 
using gates. To be specific, we shall assume that we are dealing with singlets, for 
which the system is first allowed to relax into its ground-state at low-temperature and 
zero magnetic field (or constant field in which an integral number of flux quanta pass 
through the dot). The single cell may thus be described by only two coherent singlet 
states, with charge ordering as shown in figure \ref{fig2} and effective Hamiltonian given by 
equation (\ref{Hsing}) or, equivalently, the pseudo-spin Hamiltonian, equation (\ref{spinH}), with 
$\gamma=2\Delta\cos(2\phi)$. We shall confine 
ourselves to this singlet manifold for the remainder of the paper, though it is 
straightforward to extend the results to the spin-polarised triplet states using the 
effective Hamiltonian given by equation (\ref{Htrip}). These states may be manipulated using 
the electrostatic gates in the four corners. For example, we may initialise 
the `0' state by applying a large positive bias to corner gates 1 and 3 and allowing 
the system to relax to its ground state. Removing these potentials at $t=0$
will, according to equation (\ref{Hsing}) (with $\cos2\phi=1$), cause the system to oscillate between 
the states $|0\rangle$ and $|1\rangle$ with a period $T=\pi/2\Delta$. This follows 
directly by expressing the initial state in terms of eigenvectors and applying the 
time evolution operator as follows.

\begin{eqnarray}
\label{Tdev}
 |\psi(t)\rangle = & e^{-iHt} |0\rangle & = e^{-iHt} \frac{|+\rangle + |-\rangle}{\sqrt{2}} \\ \nonumber
		    &		        & =  \frac{e^{-iE_+t} |+\rangle + e^{-iE_-t}|-\rangle}{\sqrt{2}} \\ \nonumber
		    &			& = e^{-iE_0t}(\cos 2\Delta t |0\rangle + i \sin 2\Delta t |1\rangle) 
\end{eqnarray}
where we have used
\begin{equation}
H  = \left[ {\begin{array}{*{20}c}
   {E_0}     & {2\Delta}  \\
   {2\Delta} & {E_0}      \\
\end{array}} \right]
\end{equation}
giving $E_\pm = E_0 \pm 2 \Delta$ and $|\pm\rangle = \frac{ |0\rangle + |1\rangle}{\sqrt{2}}$.

Thus, after time $\pi/8\Delta$ a Hadamard transformation has taken place, with a NOT operation at 
time $\pi/4\Delta$. These timescales are typically of order a picosecond, e.g. for a 100nm dot in 
GaAs, T=0.6ps. The 2-state time translations given by equation (\ref{Tdev}) are not 
completely general since they are restricted to a plane through the centre of the 
Bloch sphere \cite{Nielson00}. However, the most general transformation, of the form:
\begin{equation}
|\psi (t)\rangle  = e^{i\alpha (t)} \left[ {\cos \theta (t)|0\rangle  + e^{i\phi (t)} \sin \theta (t)|1\rangle } \right]
\label{gen}
\end{equation}				                     
may be achieved by applying a second rotation (time evolution) for which 
the Hamiltonian is changed by the application of a gate voltage to two diagonally 
opposite corners of the dot. This will change the energy of one of the base states 
in a predictable way. Neglecting the small change in the tunnelling matrix element $\Delta$, 
the new Hamiltonian takes the more general form of equation (\ref{Hsing}) with 
$E_0 \neq E_1$. When the energy difference between these diagonal matrix elements 
is much greater than $\Delta$, the eigenstates of the new Hamiltonian are approximately 
$|0\rangle$ and $|1\rangle$ and hence after a further time $\tau$ we get,
\begin{eqnarray}
 |\psi(t+\tau)\rangle = & e^{-iHt} |0\rangle & = e^{-it\tau} e^{-iE_0t} (\cos 2\Delta t |0\rangle + i \sin 2\Delta t |1\rangle) \\ \nonumber
		    &		        & \approx  e^{-iE_0(t+\tau)} (\cos 2\Delta t |0\rangle + i e^{-i E_g\tau} \sin 2\Delta t |1\rangle)
\end{eqnarray}
where $E_g=E_1-E_0 \approx e V_g$, and $V_g$ is the gate potential.
This clearly has the general form (\ref{gen}). It is straightforward to further generalise 
this procedure to any sequence of transformations in which gate voltages are changed in a stepwise fashion, 
\textit{including} corrections to take into account the change in off-diagonal 
matrix elements of the Hamiltonian. In practise it may be advantageous to use a 
magnetic field to control the latter since they have little effect on the diagonal 
elements. For example,  if we again apply a gate potential to the system in the state 
$|\psi(t)\rangle$ given by equation (\ref{Tdev}), and simultaneously a magnetic field 
which produces half a flux quantum through the dot (thereby switching off the off
diagonal matrix elements), we may again  generate the general transformation (\ref{gen}), 
but with the advantage that the gate voltage need not be large. It may thus be 
`tuned' to give the optimum switching time.

\section{Two-qubit gates}

     Two qubit transformations are performed using the coulomb interaction, i.e. 
the last term in the pseudo-spin Hamiltonian (\ref{effspinH}). A disadvantage of using the coulomb 
interaction for inter-qubit interactions is that it is difficult to control. 
DiVincenzo and coworkers circumvent this problem by modulating the exchange interaction 
between (real) spins on adjacent quantum dots by changing the potential barrier 
between them. In reference \onlinecite{Toth01} three cells (qubits) were considered as a means of producing 
entanglement via the direct coulomb interaction. Here we consider a simpler scheme in 
which the effect of the coulomb interaction between two cells is controlled rather 
than its absolute value. The latter may only be changed by dynamical screening between 
cells, the possibility of which we consider briefly in the final section. In order for 
the coulomb interaction to be effective in producing entanglement between adjacent 
cells, we need to initialise the cells in such a way that a simple product state, 
degenerate or nearly so with some other product state, evolves into a linear combination 
of the two product states which cannot be factored in any basis. 
A potential candidate would be to start with one of the simple product states 
$|0,0\rangle$, $|0,1\rangle$, $|1,0\rangle$ or $|1,1\rangle$, and observe the evolution. 
However, although this evolution does eventually produce full entanglement, it is dominated 
by single-qubit rotations (since $\gamma>>v$ in equation (\ref{effspinH}) with $N=2$), 
which produce rapid oscillations in the entanglement with time.
We can in principle switch off these single-bit rotations with a magnetic field, for which 
$\gamma=0$ when a half-integral flux passes through the dot. However, this would also 
not lead to entanglement since the inter-dot Coulomb interaction is diagonal in the $|0,1\rangle$
basis and the states initial state would remain a simple product. Indeed, we may choose 
such simple product states in the 'computational' basis as initial states prior to pure single qubit rotations 
in order to ensure that the states remain unentangled.
Entanglement may be produced by initialising the two-qubit state as a simple product of \textit{energy} 
eigenstates. These are obtained by diagonalising the single-qubit Hamiltonian, e.g. 
equation (\ref{spinH}) with $\varepsilon=0$ and the constant energy term dropped, 
giving $E_\pm= \pm \gamma$ and $|\pm\rangle = (|0\rangle \pm |1\rangle)/\sqrt{2}$. 
Starting in the product state $|+,-\rangle$, time evolution will mix in the state $|-,+\rangle$ via the 
coulomb interaction, without which the state would not change. Note that \textit{only} 
the $|-,+\rangle$ state is mixed in since the Hamiltonian only couples to this state. 
This is analogous to the case of two electrons on the 
surface of liquid helium, for which the computational basis is given by the ground
$(|\alpha\rangle)$ and first excited $(|\beta\rangle)$ states of each electron in a 1D Coulomb potential above
the helium surface \cite{Platzman99}. In this system, with no applied fields $|\alpha,\beta\rangle$
and $|\beta,\alpha\rangle$ are degenerate, but this degeneracy can be broken by 
a Stark shift on one of the qubits. The (weak) Coulomb interaction between adjacent qubits 
then splits the degeneracy, with eigenstates given by the symmetric and anti-symmetric 
superpositions of $|\alpha,\beta\rangle$ and $|\beta,\alpha\rangle$.  Starting with either 
$|\alpha,\beta\rangle$ or $|\beta,\alpha\rangle$ away 
from degeneracy (with the Stark shift dominating the coulomb splitting), and turning 
off the Stark shift for a chosen time, can in principle produce maximal entanglement 
(a ROOT SWAP gate) with good fidelity \cite{Lea00}. In our quantum dot system, the state $|-\rangle$ 
and $|+\rangle$ play the role of $|\alpha\rangle$ and $|\beta\rangle$ in the helium system.

\begin{figure}
\includegraphics[width=5in]{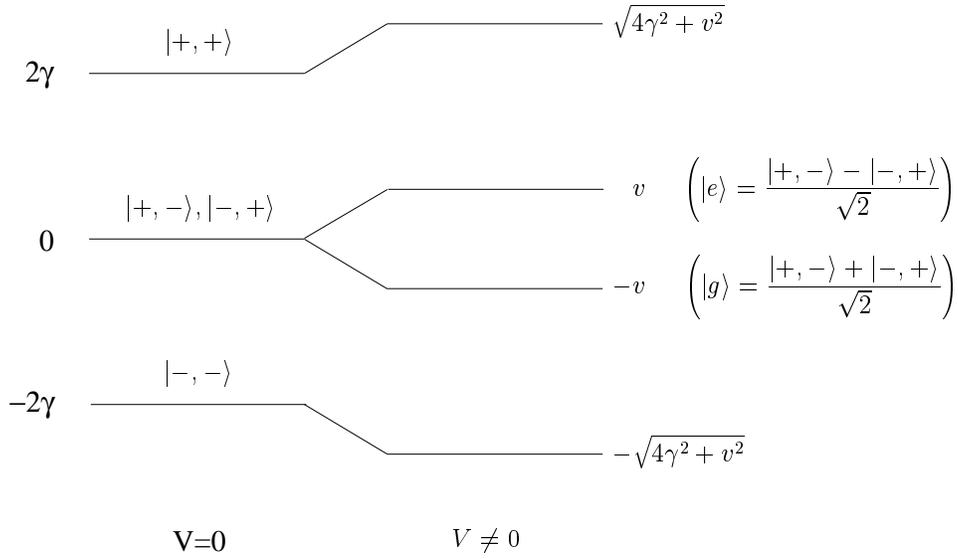}
\caption{\label{fig5}Eigensolutions of the two-qubit quantum-dot Hamiltonian with weak
coulomb repulsion between cells.}
\end{figure}

\begin{figure}
\includegraphics[height=4in]{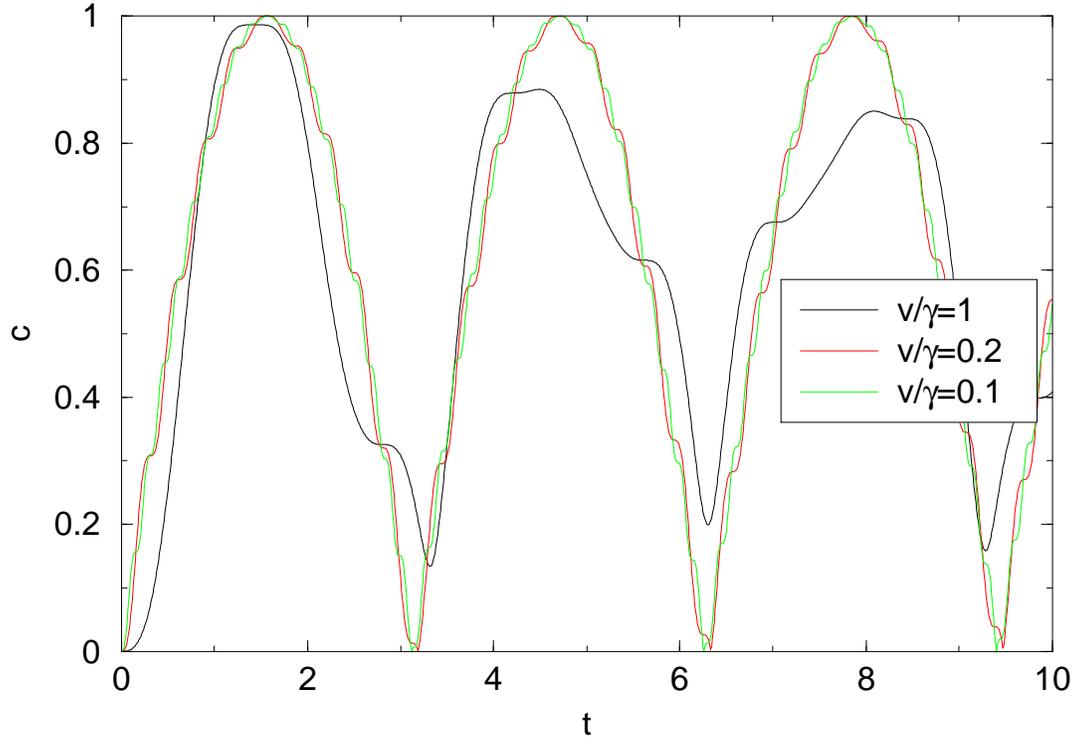}
\caption{\label{fig6}Concurrence for starting state $|0,0\rangle$ and $v/\gamma$ = 1, 0.2 and 0.2, 
showing oscillations on a timescale $h/\gamma$ before reaching full entanglement.}
\end{figure}

In the cell-diagonal energy basis, the two-qubit hamiltonian matrix is

\begin{equation}
{\bf{H}} = \left( {\begin{array}{*{20}c}
   -2\gamma & 0 & 0 & -v \\
   0 & 0 & -v & 0  \\
   0 & -v & 0 & 0  \\
   -v & 0 & 0 & 2\gamma  \\
\end{array}} \right)
\end{equation}
where the rows and columns are ordered in ascending single-cell energy. The 
eigensolutions are shown in figure \ref{fig5}.
It follows directly that the state vector at time $t$ is
\begin{eqnarray}
|\psi (t)\rangle  & = & e^{ - iHt} | +,- \rangle  = e^{ - ivt} |g\rangle  + e^{ivt} |e\rangle  \\ \nonumber
                  & = & \cos vt|+,- \rangle  - i\sin vt|-,+ \rangle  
\label{dev2}
\end{eqnarray}
which is clearly fully entangled for $t=(2n+1)\pi/4v, n=0,1,2\ldots$.
The degree of entanglement is not so apparent in the original 
$|0\rangle$, $|1\rangle$ basis. However, we may obtain an absolute measure of entanglement by computing 
the so-called concurrence \cite{Wooters98} which is basis independent,
\begin{equation}
c(t) = 2 | \langle a,a | \psi(t) \rangle \langle b,b | \psi(t) \rangle - 
           \langle a,b | \psi(t) \rangle \langle b,a | \psi(t) \rangle |
\end{equation}
where $|a\rangle$ and $|b\rangle$ are any orthonormal single-qubit basis states. 
Choosing these to be the energy eigenstates we see directly from (\ref{dev2}) that
\begin{equation}
c(t) = |\sin 2vt|
\label{concur}
\end{equation}
showing that $c(t)=1$ again for $t=(2n+1)\pi/4v, n=0,1,2\ldots$.
We note that the entanglement depends only on $v$ and is 
independent of $\gamma$, the energy parameter which controls single-qubit  rotations.
By contrast, a plot of $c(t)$ starting with a product state in the 'computational' basis,
shows rapid oscillations on a timescale of $h/\gamma$, as shown in figure \ref{fig6}.

Two further issues arise when considering the above procedure for producing entangled states. 
The first is how to prepare the initial state, $|+,-\rangle$, and secondly how to preserve an 
entangled state once it has been produced, bearing in mind that the degree of entanglement 
oscillates with time. A further, and related, potential difficulty is being able to perform 
single-qubit operations for which the effect of the intra-cell coulomb interaction is made 
negligible.  Clearly the most direct, and probably the most difficult, way to do this is by 
selectively switching off the coulomb interaction between cells (see next section). An 
indirect way is to lift the degeneracy between the states $|+,-\rangle$ and $|-,+\rangle$ whilst the single
qubit operations are being performed. For the Helium system
it is proposed to implement this through application of an electric field to one electron 
creating a Stark shift \cite{Lea00}. For the quantum dots a magnetic field is applied to one dot 
but not the other, thereby changing the value of the tunnelling parameter $\gamma$ on one 
of the dots. For example, if the magnetic flux through one of the dots is half a flux quantum 
the corresponding coupling is zero and the $|+\rangle$ and  $|-\rangle$ states for that dot become degenerate. 
Hence the states $|+,-\rangle$ and $|-,+\rangle$ now differ in energy by $2\gamma$, where
$\gamma$  is the tunnelling parameter on the other dot. The eigensolutions of the two-cell Hamiltonian 
are then $E=\pm \varepsilon$ and $|g\rangle = \cos \theta |+,-\rangle + \sin \theta |-,+\rangle$ and
$|e\rangle = -\sin \theta |+,-\rangle + \cos \theta |-,+\rangle$, where $\varepsilon = \sqrt{\gamma^2 + v^2}$
and $\tan \theta = \frac{v}{\gamma + \sqrt{\gamma^2 + v^2}}$. If we again start in the state 
$|+,-\rangle$ at $t=0$, the state at time $t$ becomes
\begin{equation}
|\psi (t)\rangle  = e^{i\varepsilon t} \sin \theta |g\rangle  + \cos \theta e^{ - i\varepsilon t} |e\rangle  = \left[ {e^{i\varepsilon t} \cos ^2 \theta  + e^{ - i\varepsilon t} \sin ^2 \theta } \right]| + , - \rangle  + i\sin 2\theta \sin \varepsilon t| - , + \rangle
\label{eq16}
\end{equation}
Hence, the probability of finding the (other) state $|-,+\rangle$ at time $t$ is 
$|\langle -,+ |\psi(t)\rangle |^2 = \sin^2 2 \theta \cos^2 \varepsilon t$
which has a  maximum value of $\sin^2 2 \theta \approx (v/\gamma)^2$ 
when $v/\gamma$ is small. This should be compared with the 
previous case for which the states 
and the maximum value of $|\langle -,+ | \psi(t) \rangle |^2$ is unity. 

\begin{figure}
\includegraphics[height=4in]{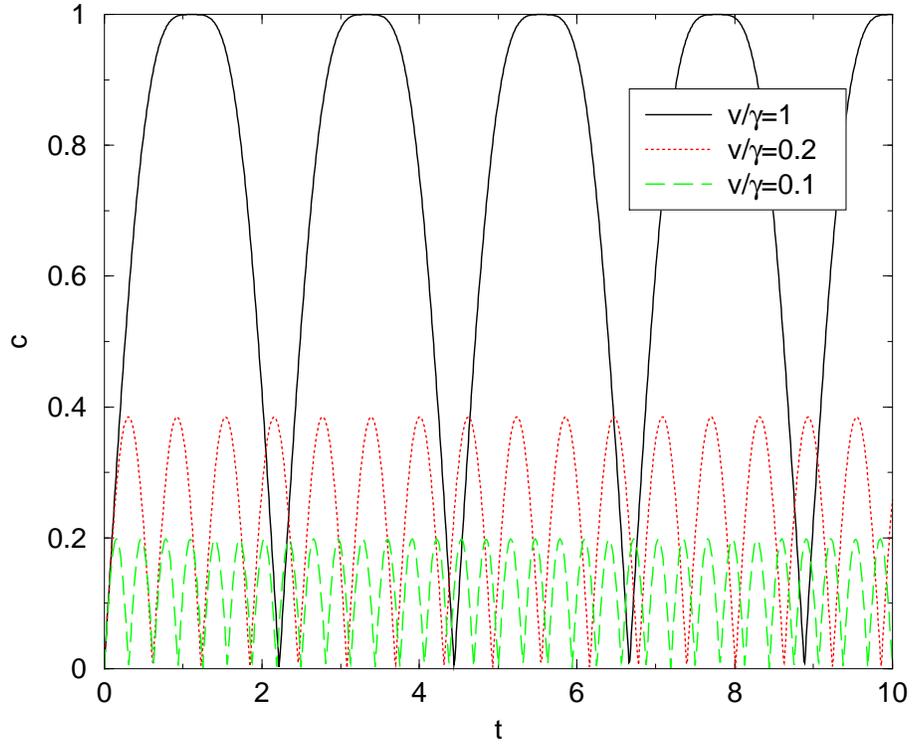}
\caption{\label{fig7}Time dependence of concurrence for an initial state $|+,-\rangle$ in 
which $\gamma=0$ on one dot.}
\end{figure}

An alternative measure of the effectiveness of `detuning' in preserving a state is the 
concurrence. This is zero for the state $|+,-\rangle$ at $t=0$ and  
$2 |\sin 2\theta \sin \varepsilon t \sqrt{1-\sin^2 2 \theta \sin^2 \varepsilon t}|$   at later times when 
the system is in the state (\ref{eq16}). This is plotted in figure \ref{fig7} with $\gamma=0$ on the one 
dot and $v/\gamma=0.1, 0.2$ and $1$ on the other.
We see that the maximum concurrence decreases 
rapidly as the intra-dot tunnelling increases relative to the inter-dot coulomb interaction, 
with $c_{max}(t)\approx 2 \sin 2\theta \approx 2 v/\gamma = 0.2$ for $v/\gamma=0.1$.
Note also that when $v/\gamma$ is small, the concurrence oscillates with angular frequency 
$\varepsilon\approx\gamma$. This should be compared with the case when $\gamma$ is the same 
on both dots, for which $c(t)$ oscillates between 0 and 1 at the much lower frequency $2v$ 
(equation \ref{concur}). We note that such detuning of the initial state in order to remove its 
degeneracy with the state $|-,+\rangle$ may also be done by applying positive gate voltages to 
all four corner gates on one dot. This will change the confinement potential profile 
such as to again reduce the tunnelling parameter $\gamma$. 

Although the above de-tuning procedure will significantly reduce the maximum concurrence 
caused by the Coulomb interaction, this may still give rise to systematic errors for 
single-qubit transformations that are not sufficiently small to be quantum error corrected. 
However, we have shown that the entangling effect of the Coulomb interaction may be eliminated 
completely by starting with product states that are diagonal in the 'computational' basis 
(i.e. the states $|0,0\rangle$, $|0,1\rangle$, $|1,0\rangle$ or $|1,1\rangle$) and setting 
$\gamma=0$ on one of the dots (or all dots apart from one in an array of qubits). Since the 
Coulomb interaction is diagonal in this basis, we might expect that it would be ineffective 
in producing entanglement since there are no single-qubit rotations on the dots for which 
$\gamma=0$.  This is indeed the case as may be proved by expressing the initial state in 
the basis of exact two-qubit eigenvectors, propagating with the time evolution operators, 
and re-expressing the result in product basis states, from which the concurrence is calculated 
and shown to be zero by exact cancellation.  A related situation exists in the proposed 
scenario for Josephson junction qubits \cite{Schon99,Schon00}, where it is not possible 
to do simultaneous one-qubit operations on adjacent qubits without turning on the 
(controllable) coupling between these qubits.

\begin{figure}
\includegraphics[height=4in]{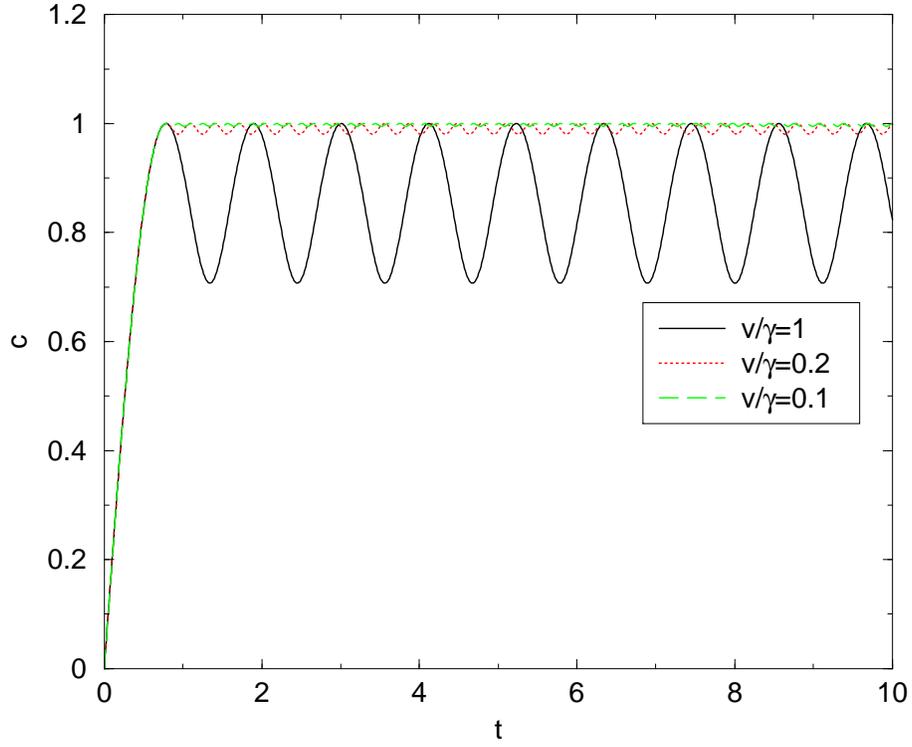}
\caption{\label{fig8}Time dependence of concurrence with an initial state $|+,-\rangle$
for which $\gamma$ is switched off on one dot only at $t=\pi/4v$. }
\end{figure}

The concurrence may also be preserved by detuning a state that is maximally entangled. 
For example, we have shown that a fully entangled state may be produced at time 
$t=\pi/4v$ (see equation \ref{dev2}). If at this time we change the Hamiltonian 
by switching off the tunnelling process in one dot (as described above), then the concurrence is,
\begin{eqnarray}
c(t) & = & |\sin 2 v t|    \hspace{2.3in} (t \leq \pi/4v) \\ \nonumber
     & = & \sqrt{1-\sin^2 2 \theta \sin^2 2 \varepsilon (t - \pi/4v)} \hspace{0.8in}  (t \geq \pi/4v)
\end{eqnarray}
This is plotted in figure \ref{fig8}, again for $v/\gamma=0.1, 0.2$ and $1$, and we see 
that $c(t)$ remains close to unity when $v/\gamma$ is small.
In fact, preservation of an entangled state is more effective than preservation of an unentangled 
state by this method (compare figure \ref{fig7}). Even the case $\gamma=v$ has a minimum 
entanglement of $c_{min} = 1/\sqrt{2} = 0.7071$ and rapidly approaches unity with 
decreasing $v/\gamma$.
For small $v/\gamma$ the correction $\sim (v/\gamma)^2$ giving maximum deviations in concurrence 
of order a percent for realistic parameter values. Whilst this may still not be sufficiently 
small to allow full quantum error correction, it is well within the error necessary to 
unequivocally observe entanglement experimentally. 

With the qubits essentially decoupled, we may prepare the initial state $|+,-\rangle$ by 
single-qubit rotations. A novel way to do this is to first allow the two qubits to 
relax into their ground states, when a magnetic field  is applied to one dot such 
that the magnetic flux through the dot is one flux quantum. Removing this magnetic 
field at $t=0$ on the first dot will give the desired state since this exchanges 
the ground and excited states \cite{Creffield00}. 

We have discussed here the realization of a ROOT SWAP gate, which is a suitable 
universal two-qubit gate \cite{DiVincenzo95,Lloyd95,Barenco95}. It is well known 
how to convert such a gate into other two-qubit gates, such as a controlled phase 
 flip or a CNOT, using single qubit rotations \cite{Burkard99}. In any actual computation, 
the most suitable two-qubit gate(s) would depend upon the algorithm, also taking into 
account physical optimization \cite{Burkard99}. For now, we remark that in the initial 
stages of investigation of new candidates for quantum computation, the most important 
two-qubit issue is the capability to produce entanglement. Thus the simplest gate to 
consider is that which requires no (or the minimum number of) additional single qubit 
gates. Experiments to investigate entanglement in dot systems such as those proposed 
here will therefore probably focus first on ROOT SWAP.

\section{Summary and Discussion}

In this paper we have analysed the potential of two electrons in square semiconductor 
quantum dots as a basis for a scalable array of quantum computing elements. As with all 
few-electron quantum dots in the strong correlation regime,  there exists an isolated 
low-lying multiplet of states which has the potential to form a controllable finite-state 
system, or `qumit'. For two electrons, this is a two-state, i.e qubit, system for the 
orbital degrees of freedom. This requires that the spin-state (singlet or triplet) does 
not change during the computation period between initialisation and measurement. It 
has been argued elsewhere \cite{DiVincenzo99_1} that this is feasible due to the long lifetime of spin 
states in semiconductors, though it is by no means certain that these will be sufficiently 
long for practical implementation of a spin-based quantum computer. In addition to this, 
we require that the orbital states be sufficiently long-lived for meaningful computations 
to be made. This is also uncertain at present though various schemes can be envisaged 
which would reduce the inelastic scattering of electrons by the relevant acoustic phonons 
of energy equal to the energy splitting of the ground multiplet states. Note that at 
low-temperatures the main process would be spontaneous phonon emission and it had been suggested that 
this may be inhibited by device geometry and material choice at heterojunction interfaces \cite{Khaetskii00}. 
Indeed, very recent theoretical work has demonstrated this quantitatively \cite{Debald02}. A 
further possibility would be a molecular system for which the relevant low-lying electronic 
energy splitting would be chosen so that they were not resonant with the molecular 
vibrational modes. In order that the electrons remain strongly correlated in such a 
small molecular system, they would have to be screened by filled outer shells as in 
d or f electron systems \cite{Lent00}.

Initialisation of our proposed system is performed by corner gates, which force each 
qubit into one of the two computational-basis states. Subsequent single-qubit operations 
(rotations) are then performed by removing the voltages on the corner gates and allowing 
the system to develop ballistically in time. Further control may be achieved  during 
the computation phase by the application of a magnetic field, which changes the 
magnitude of the tunnelling energy parameter, and by further application of corner 
gate voltages. In this way we have shown that any single-qubit transformation may be performed.

We have shown how two-qubit transformations may be performed via the coulomb interaction 
between qubits provided this is small compared with the intra-qubit tunnelling energy. 
A potential problem is the control of the coulomb interaction itself. This is achieved 
by appropriate choice of initial states and intermediate states which are `tuned' by 
changing the tunnelling parameter on one of the qubit-dots. In this way, entanglement 
is maximised  via an initial state in which one dot is in its ground state and the other 
in its excited state. When this has been achieved, an entangled state is preserved by 
changing the tunnelling parameter in one dot only, via a magnetic field.  A more direct 
control of the Coulomb interaction would be to physically change its magnitude during 
computation. This may be achieved in principle by introducing screening charge between 
the qubits.  Control of this charge could be via surface gates or by controlling 
supercurrents, which are driven normal by exceeding their critical value. Although such 
schemes would be technologically challenging, they cannot be ruled out on fundamental 
grounds and may ultimately be a practical way forward in directly controlling the 
coulomb interaction. 

Finally, we consider readout, i.e. the means of measuring the state after computation.  
For the spin system proposed by DiVincenzo and coworkers, a suggested method was to use 
spin-dependent tunnel barriers in which an electron of the appropriate spin would tunnel 
into a quantum confined region where it would be detected by a single-electron transistor 
\cite{DiVincenzo99}. A similar scheme is envisaged here but with the added simplicity that the tunnel 
barrier would not be spin dependent. It would be located at one or more corners of the 
quantum dot (one is sufficient in principle but more may be desirable to reduce errors). 
When the qubit state is measured, the tunnel barrier is lowered and if an electron is 
in that corner of the dot it will tunnel into a quantum-confined region and be detected 
by the single-electron transistor. This uniquely defines the measured state. After 
measurement the system may be `reset' using appropriate gate bias voltages.

\begin{figure}
\includegraphics[height=4in]{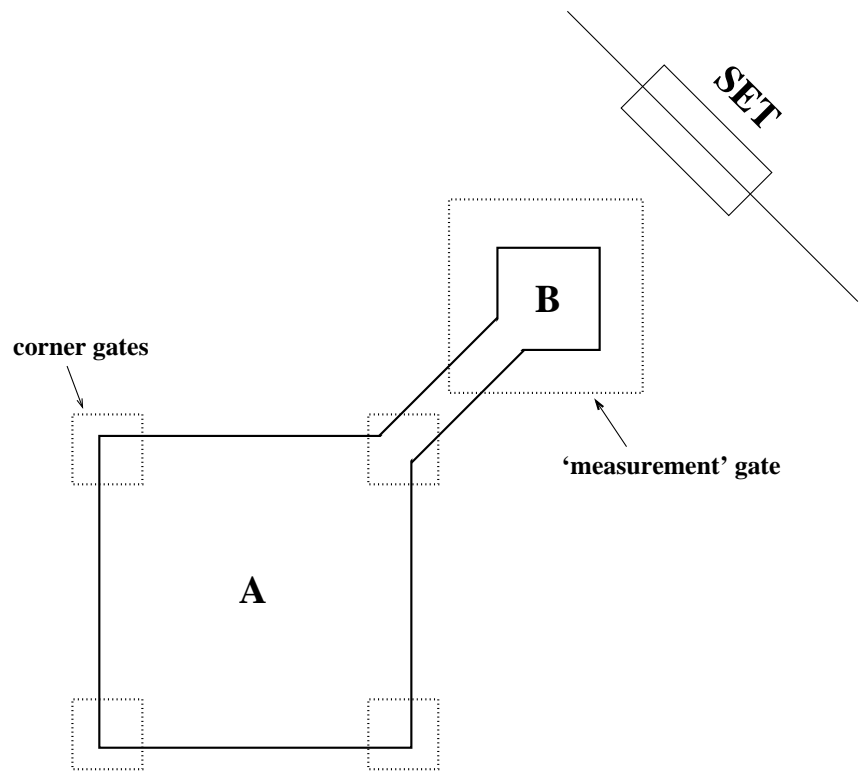}
\caption{\label{fig9}Schematic diagram showing arrangment for final-state measurement.}
\end{figure}

Note that in order for a coherent time-development of the qubit to be negligible 
during the measurement process, the tunnel barrier must be lowered and raised on a 
timescale which exceeds the electron tunneling time but is sufficiently short for 
intra-dot time-development (e.g. single-qubit rotations) to be small, ideally negligible. 
This may be achieved in principle by the arrangement of gates shown schematically in figure \ref{fig9}.
The square quantum dot (A)  is extended at one corner by a quantum wire leading to a further 
quantum dot (B). This is essentially two very weakly coupled quantum dots for which the two 
electrons are normally located in the larger dot, A. As far as the quantum computation is concerned, 
the dots are decoupled and the states are controlled by corner gates as described earlier.  
The measurement is performed by lowering the electron potential energy in dot B via a positively 
biased gate (shown dotted). If an electron is in the corner of dot A, which is connected to dot B, 
then it will be transferred to the latter. Subsequently raising the electron potential energy of 
dot B to its 'normal' voltage will trap the electron in dot B where it can be later measured by 
a single-electron transistor. It should be realized that the process of transferring the electron 
from dot A to dot B does not necessarily destroy the coherence of the full quantum state involving 
both dots. If the initial single-qubit state is a superposition of states $|0\rangle$ and $|1\rangle$ 
then the final state after the possible transfer of an electron will also be a superposition of 
states in which an electron is either in dot B or not. However, since the two dots are essentially 
decoupled, this superposition does not change with time, unlike the 'quantum computation' states 
which continually oscillate between the base states. It is only when the presence (or not) of an 
electron on dot B is measured by the single-electron transistor that the wavefunction 'collapses' 
giving a probability that is essentially $|\langle 0|\psi\rangle|^2$.

\begin{acknowledgments}
The authors would like to acknowledge helpful discussions with J. Annett, C. Creffield , B. Gyorffy,
C. J. Lambert, I. A. Larkin, P. Meeson and W.J. Munro. This work was supported in part by the
UK Ministry of Defence (corporate research programme).
\end{acknowledgments}

\newpage 

\bibliography{Qubit}   

\newpage

\end{document}